\documentclass[prb,twocolumn,superscriptaddress,showpacs]{revtex4}
\usepackage{graphicx}
\usepackage{amsmath}
\usepackage{color}
\begin{document}

\title{Enhancement of spin propagation due to interlayer exciton condensation}
\author{Louk Rademaker}
\email{rademaker@lorentz.leidenuniv.nl}
\affiliation{Institute-Lorentz for Theoretical Physics, Leiden University, PO Box 9506, Leiden, The Netherlands}
\author{Jeroen van den Brink}
\affiliation{Institute for Theoretical Solid State Physics, IFW Dresden, 01171 Dresden, Germany}
\affiliation{Department of Physics, TU Dresden, D-01062 Dresden, Germany}
\author{Hans Hilgenkamp}
\affiliation{Institute-Lorentz for Theoretical Physics, Leiden University, PO Box 9506, Leiden, The Netherlands}
\affiliation{Faculty of Science and Technology and MESA+ Institute for Nanotechnology, University of Twente, P.O. Box 217, 7500 AE Enschede, The Netherlands}
\author{Jan Zaanen}
\affiliation{Institute-Lorentz for Theoretical Physics, Leiden University, PO Box 9506, Leiden, The Netherlands}
\date{\today}

\begin{abstract}
We show that an interlayer exciton condensate doped into a strongly correlated Mott insulator exhibits a remarkable enhancement of the bandwidth of the magnetic excitations (triplons). This triplon is visible in the dynamical magnetic susceptibility and can be measured using resonant inelastic X-ray scattering. The bandwidth of the triplon scales with the exciton superfluid density, but only in the limit of strong correlations. As such the triplon bandwidth acts as a probe of exciton-spin interactions in the condensate.
\end{abstract}


\maketitle

\begin{figure}
	\includegraphics[width=\columnwidth]{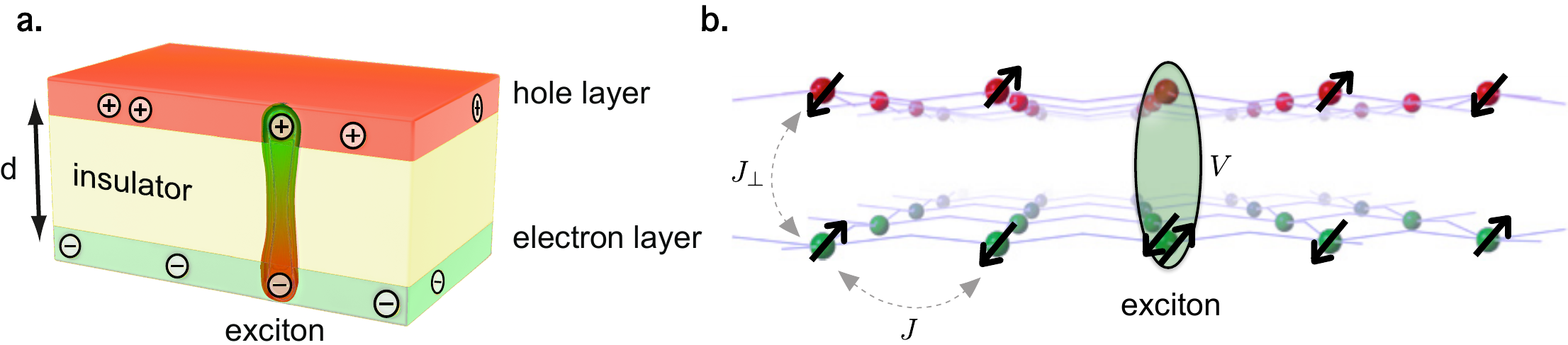}
	\caption{\label{FigEC}\textbf{a.} Interlayer excitons arise when one stacks an electron and a hole layer in a heterostructure, separated by an insulator. The Coulomb attraction between the negative electrons and positive holes creates bound electron-hole pairs, commonly referred to as `interlayer excitons'. \textbf{b.} In the presence of strong electron-electron interactions the electrons localize and only the spin degree of freedom remains. An exciton can now be viewed as the bound state of a double occupied and an empty site (a doublon-holon pair). The remaining electrons are not paired into excitons and contribute to the magnetic behavior.}
\end{figure}

Shortly after the BCS theory explained superconductivity in terms of electron-electron pairing\cite{Bardeen:1957p5048}, it was proposed that similar pairing of electrons and holes might occur\cite{Blatt:1962p4000,Keldysh:1968p4790,Moskalenko:2000p4767}. While the binding of an electron and a hole into an exciton has the advantage of the much stronger Coulomb attraction, the possible recombination and annihilation of an exciton prevents the practical realization of a so-called exciton condensate. However, if one is able to spatially separate the electrons and holes in distinct layers, as shown in figure \ref{FigEC}a, annihilation can be suppressed\cite{Shevchenko:1976p4950,Lozovik:1976p4951} and an equilibrium density of excitons can be created. Over the last decade such bilayer systems became experimentally within reach, first in quantum Hall bilayers\cite{Eisenstein:2004p4770,Huang:2012p5449} and more recently in systems without magnetic field \cite{High:2012p5349}.

The interlayer exciton condensate thus obtained is in many regards similar to the Cooper pair condensate, where the phenomenon of counterflow superfluidity\cite{Su:2008p4797,Finck:2011p5284} replaces the usual electric supercurrents. The application of an in-plane magnetic field leads to a small diamagnetic response\cite{Shevchenko:1976p4950,Balatsky:2004p3613,Rademaker:2011p5169,Eastham:2012p5445} which is the analogue of the Meissner effect.

\begin{figure*}
	\includegraphics[width=\textwidth]{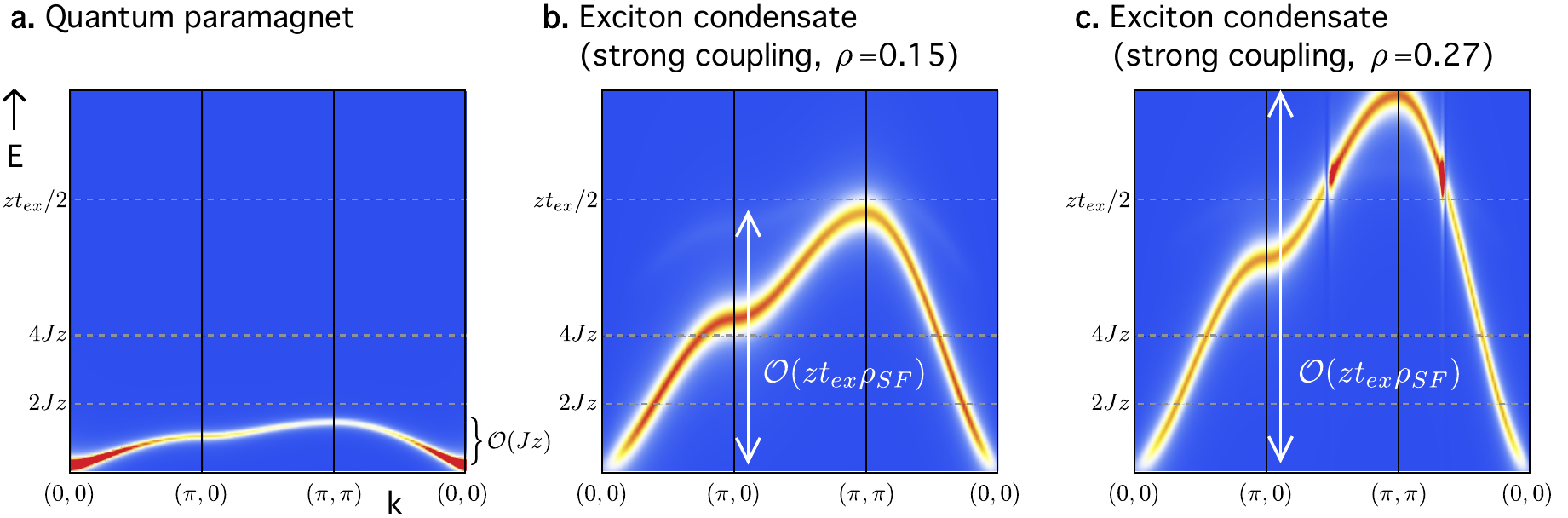}
	\caption{\label{FigStrongCLayerResults}
	The absorptive part of the dynamical magnetic susceptibility $\chi''(\mathbf{q},\omega)$ in a Mott insulating bilayer (a) doped to become an exciton condensate (b,c). 
	\textbf{a:} The spectrum of a Mott insulating bilayer with the same gap as the exciton condensates of figure b and c. The gap determined by the interlayer Heisenberg coupling $J_\perp$ is taken to be so small that it is barely visible. The bandwidth of the triplon mode is of the order $Jz$.
	\textbf{b.} In the presence of the exciton condensate, the magnetic excitation spectrum consists of propagating triplets with a gap in their spectrum as in figure a.
	Instead of the small $\mathcal{O}(Jz)$ bandwidth, the triplet has now an enhanced bandwidth $\mathcal{O}(zt_{ex} \rho_{SF})$, proportional to the superfluid density. The enhancement can be understood as a result of the condensation of the interlayer excitons, whereby the exciton-spin interaction is transformed into a mechanism that promotes the propagation of the free triplets, see equations (\ref{ExSpinExch}) and (\ref{ESCondensation}). 
	This result is computed using a linear spin wave approximation, using model parameters $t_{ex} = 2, J = 0.125, \alpha = 0.04$ and $\rho=0.15$.
	\textbf{c.} The same result as in b, but now with a higher exciton density $\rho=0.27$. The triplet mode bandwidth is seen to scale with the exciton superfluid density.}
\end{figure*}

However, the interplay between excitons and spins turns out to be nontrivial in an exciton condensate doped into a Mott insulating bilayer\cite{Ribeiro2006,Millis:2010p5231,Rademaker2012EPL,Rademaker2012arXiv}. In the absence of excitons such a bilayer orders antiferromagnetically, but in the presence of an exciton condensate the noncondensed electrons (see figure \ref{FigEC}b) form a quantum paramagnet, which has as elementary magnetic excitations the triplet modes (triplons)\cite{Sachdev:Book}. One expects that the bandwidth of the triplons is proportional to the superexchange energy $J$. However, as we will show in this Rapid Communication, interlayer exciton condensation leads to a drastic increase of the triplon bandwidth. This enhancement is rooted in the triplons 'borrowing' itineracy from the exciton condensate. The resulting bandwidth turns out to be proportional to the superfluid density, as is shown in figure \ref{FigStrongCLayerResults}. In principle, this enhancement can be detected by measurements of the dynamical magnetic susceptibility. It appears unlikely that such bilayer exciton systems can be manufactured in bulk form which is required for neutron scattering, while there is a real potential to grow these using thin layer techniques. Therefore the detection of the triplon bandwidth enhancement forms a realistic challenge for resonant inelastic X-ray scattering (RIXS)\cite{Ament:2010p5208} measurements with its claimed sensitivity for interface physics\cite{Dean2012}. 


Experimentally, various materials have been realized that could exhibit enhanced spin propagation. For example, multilayers have been fabricated using $p$- and $n$-type cuprates\cite{Imada:1998p2790,Lee:2006p1688} such as YBa$_2$Cu$_3$O$_{7-x}$ or La$_{2-x}$Sr$_x$CuO$_4$ and Nd$_{2-x}$Ce$_x$CuO$_4$. For sufficiently strong doping these materials both turn superconducting, and indeed also superconducting $p-n$ heterostructures have been realized\cite{Mao93}. For the investigation of exciton properties in a magnetic background one would prefer to stay with lower doping levels, which is currently experimentally pursued in our laboratories\cite{Hoek13}.

Another interesting material to consider in this respect is the self-doped cuprate Ba$_2$Ca$_3$Cu$_4$O$_8$F$_2$\cite{Iyo02}, which from formal valency considerations should be a Mott insulator, but in which of the four CuO$_2$ layers in the unit cell two turn out to be electron-doped and two hole-doped\cite{Shi07-08-09}. In these layers superconductivity (with $T_c \approx 55$ K) is found to coexist with antiferromagnetism (with $T_N \approx 100$ K). Though no direct signatures of exciton physics have been detected in this material, magnetic excitation measurements might elucidate similar interaction effects between the Cooper pair condensate and the spins as proposed in this Rapid Communication.

Finally, we mention the novel area of interface conductance in oxide insulators, which entails intriguing prospects to realize closely coupled $p$- and $n$-type conductors. An example has been provided by Pentcheva et al.\cite{Pen10} for the case of 2 unit cells of LaAlO$_3$ and 1 unit cell of SrTiO$_3$ grown epitaxially on a TiO$_2$-terminated SrTiO$_3$ substrate. This research-area also extends to interfaces with Mott insulator compounds such as LaVO$_3$/SrTiO$_3$\cite{Hot07}.

Let us now derive our theoretical predictions, first by introducing the exciton condensate in somewhat more detail. As mentioned above, the exciton condensate is the result of the direct interlayer Coulomb attraction, in stark contrast to the retarded phonon mediated electron-electron pairing in superconductors. Consequently, the pairing mechanism is remarkably simple and in the absence of spin-orbit coupling or magnetization the excitons are a singlet pair. The resulting condensate wavefunction has the standard BCS-form,
\begin{equation}
	| \Psi \rangle = \prod_{\mathbf{k} \sigma} \left( u_\mathbf{k}
		+ v_{\mathbf{k}} c^\dagger_{\mathbf{k}1 \sigma} c_{\mathbf{k} 2 \sigma} \right) 
		| \Psi_0 \rangle
	\label{WaveFunc}
\end{equation}
where $|\Psi_0 \rangle$ is the ground state without excitons, $ c^\dagger_{\mathbf{k}1 \sigma}$ creates an electron in the first layer and $c_{\mathbf{k} 2 \sigma}$ creates a hole in the second layer with opposite spin. The order parameter can be defined independently of the spin as
\begin{equation}
	\Delta_{\mathbf{k}} = u_\mathbf{k} v_{\mathbf{k}} 
	= \langle c^\dagger_{\mathbf{k}1 \sigma} c_{\mathbf{k} 2 \sigma} \rangle.
	\label{OrderParameter}
\end{equation}
The anomalous interlayer tunneling now serves as a direct probe of the order parameter\cite{Eisenstein:2004p4770}.

The enhancement of the triplet mode, described in detail below, is an effect that only occurs in the regime of strong electron-electron interactions. In such Mott insulators electrons localize due to interactions and only their spin remains as a degree of freedom. Bilayers (figure \ref{FigEC}b) are described by the bilayer Heisenberg model\cite{Manousakis1991,Chubukov1995}
\begin{equation}
  H_J=
  	J \sum_{\langle ij \rangle,\ell}\mathbf{S}_{i\ell}\cdot\mathbf{S}_{j\ell}
  	+J_{\perp}\sum_{i}\mathbf{S}_{i1}\cdot\mathbf{S}_{i2}.
	\label{HJ}
\end{equation}
The operators $\mathbf{S}_{i\ell}$ denote the spin of a particle on site $i$ in layer $\ell$, and via the mechanism of superexchange spin excitations can propagate. The superexchange parameters $J$ are related to the bare electron hopping $t$ by second-order perturbation theory, hence $J = 2t^2/U$ and $J_\perp = 2t_\perp^2/U$ with $U$ the onsite repulsion. While in realistic Mott bilayers $J_\perp < J$, we can use the bilayer Heisenberg model also to describe a generic quantum paramagnet. For this we need to artificially put $J_\perp \gg J$, thus favoring singlet configurations on each interlayer rung. The excitation spectrum consists of propagating triplet modes, with a dispersion $\omega_\mathbf{k} = Jz \sqrt{\alpha (\alpha - \gamma_\mathbf{k})}$ where $\alpha = J_\perp / Jz$, $z$ is the lattice coordination number and $\gamma_\mathbf{k} = \frac{1}{2} (\cos k_x + \cos k_y)$. Hence the bandwidth of these triplets in the absence of exciton condensation is set by the superexchange parameter $J$. We compute the interlayer dynamical magnetic susceptibility\cite{BruusFlensberg}
\begin{equation}
	\chi_{ij} (\tau) =
		\langle T_\tau (S^-_{i 1} (\tau) - S^-_{i 2} (\tau))( S^+_{j 1}-S^+_{j2}) \rangle
	\label{TotalSusc}
\end{equation}
using the well-tested linear spin wave theory\cite{Manousakis1991,Chubukov1995}. The imaginary part $\chi''$, which describes the absorption, is in principle measurable by RIXS\cite{Ament:2010p5208} and a typical expected spectrum is shown in figure \ref{FigStrongCLayerResults}a.

As for the case of normal carriers in a doped Mott insulator, the nature of the exciton system is drastically different from what is found in uncorrelated semiconductors. The Mott insulator cannot be described by band theory, and instead electron- and hole-doping corresponds with the creation of double occupied sites (doublons) and empty sites (holons), respectively. The doublons and holons attract each other via the Coulomb attraction and can thus form doublon-holon pairs: the strong coupling limit of the exciton shown in figure \ref{FigEC}b. Since in the Mott bilayer all interactions are strong, the relevant case is to assume strong exciton binding such that excitons can be treated as local pairs and the condensation occurs in the BEC sense rather than in the weak coupling BCS sense.

To describe such a doublon-holon pair in a Mott bilayer, we can express the exciton hopping in terms of interlayer rung states: the exciton $| E \rangle$ and the four possible interlayer spin states $| s \; m \rangle$. The motion of an exciton is governed by\cite{Rademaker2012EPL,Rademaker2012arXiv}
\begin{equation}
	H_K = - t_{ex} \sum_{\langle ij \rangle} | E \rangle_j \left(  \sum_{sm} |s \; m \rangle_i \langle s  \; m |_j
			 \right) \langle E |_i.
		\label{ExcitonHop}
\end{equation}
The exciton hopping energy $t_{ex}$ can be related to the electron hopping via second-order perturbation theory, which gives $t_{ex} = t^2/V$ where $V$ equals the binding energy of an exciton.

The system describing coexistence of spins and excitons, given by equations (\ref{HJ}) and (\ref{ExcitonHop}), is equivalent to a hard-core boson system, reminiscent of attempts to describe cuprate superconductivity using only bosons such as the $SO(5)$ theory of the $t-J$ model\cite{Zhang:1997p1677}. In contrast to these theories, for the excitons in Mott bilayers the mapping onto bosonic physics is fully controlled. The ground state of the `exciton $t-J$ model' can straightforwardly be found using a $SU(5)$ coherent state. 
Elsewhere we study this in detail\cite{RademakerFuture} finding that the dynamical frustration between excitons and spins causes large parts of the phase diagram to be dominated by phase separation. As long as the exciton hopping $t$ is bigger than the exciton-exciton repulsion we find an exciton superfluid as the ground state, where the spins form interlayer singlets. Given strongly bound excitons, the critical temperature for condensation can be as high as 700 Kelvin\cite{RademakerFuture}. Note that in principle there can be sign problems\cite{Wu:2008p3149} but these drop out rigorously for this singlet ground state.

The exciton condensate wavefunction is now
\begin{equation}
	| \Psi \rangle = \prod_{i } \left( \sqrt{\rho} \, | E \rangle_i
		+ \sqrt{1-\rho} \, | 0 \; 0 \rangle_i \right) 
\end{equation}
where $|0 \; 0 \rangle$ is the interlayer singlet spin configuration and $\rho$ is the exciton density. This is equal to equation (\ref{WaveFunc}) when $\Delta_\mathbf{k}$ is independent of momentum.

Since we are dealing with hard-core bosons forming a mean field ground state, the magnetic excitation spectrum can be computed with linear spin wave theory. We employ the Heisenberg equations of motion which are decoupled exploiting the ground state expectation values\cite{Zubarev1960,Oles:2000p5088}. The resulting dynamical magnetic susceptibilities $\chi''(\mathbf{q},\omega)$ are shown in figure \ref{FigStrongCLayerResults}, for two choices of exciton density $\rho=0.15$ and $\rho=0.27$. 

These figures illustrate the central result of this Rapid Communication: compared to the undoped system (figure \ref{FigStrongCLayerResults}a) we find that the triplon bandwidth is greatly enhanced (figures \ref{FigStrongCLayerResults}b and c). The mechanism is actually similar to that in slave-boson theories\cite{Lee:2006p1688}, where four-operator products $b^\dagger b f^\dagger f$ are decoupled as $\langle b^\dagger \rangle \langle b \rangle f^\dagger f$ yielding kinetic energy for the $f$-excitations. For Mott bilayers, we can explicitly introduce Fock operators for the exciton $e^\dagger = |E \rangle \langle 0 |$ and the triplet $t^\dagger = |1m \rangle \langle 0 |$. This implies that the exciton-spin interaction term (\ref{ExcitonHop}) can be written as
\begin{equation}
	-t_{ex} \sum_{\langle ij \rangle} e_j^\dagger e_i t^\dagger_i t_j.
	\label{ExSpinExch}
\end{equation}
This is a higher order exchange term, which at first sight seems to be irrelevant for the bandwidth of the triplet. However, when the exciton condensate sets in, the operator $e^\dagger$ obtains an expectation value $\langle e^\dagger \rangle = \sqrt{\rho_{\mathrm{SF}}}$, where $\rho_{\mathrm{SF}}$ is the condensate density. Consequently this exchange term turns into an effective triplet hopping term
\begin{equation}
	-t_{ex} \rho_{\mathrm{SF}} \sum_{\langle ij \rangle} t^\dagger_i t_j
	\label{ESCondensation}.
\end{equation}
The explains why the bandwidth of the triplet excitations is increased by an amount of order $zt_{ex} \rho_{\mathrm{SF}}$.

Surely, we made the argument that this effect leads to a dramatic increase of the bandwidth, for which we have implicitly assumed that $t_{ex}$ is larger than $J$. Now the exciton hopping energy is related to the electron hopping by $t_{ex}= t^2/V$, while the spin superexchange satisfies $J=2t^2/U$ where $U$ is the onsite Coulomb repulsion. Since for obvious reasons $U > V$, we find that indeed the dominant scale controlling the triplon bandwidth is $zt_{ex} \rho_{\mathrm{SF}}$ yielding the predicted bandwidth enhancement.

\begin{figure}
	\includegraphics[width=\columnwidth]{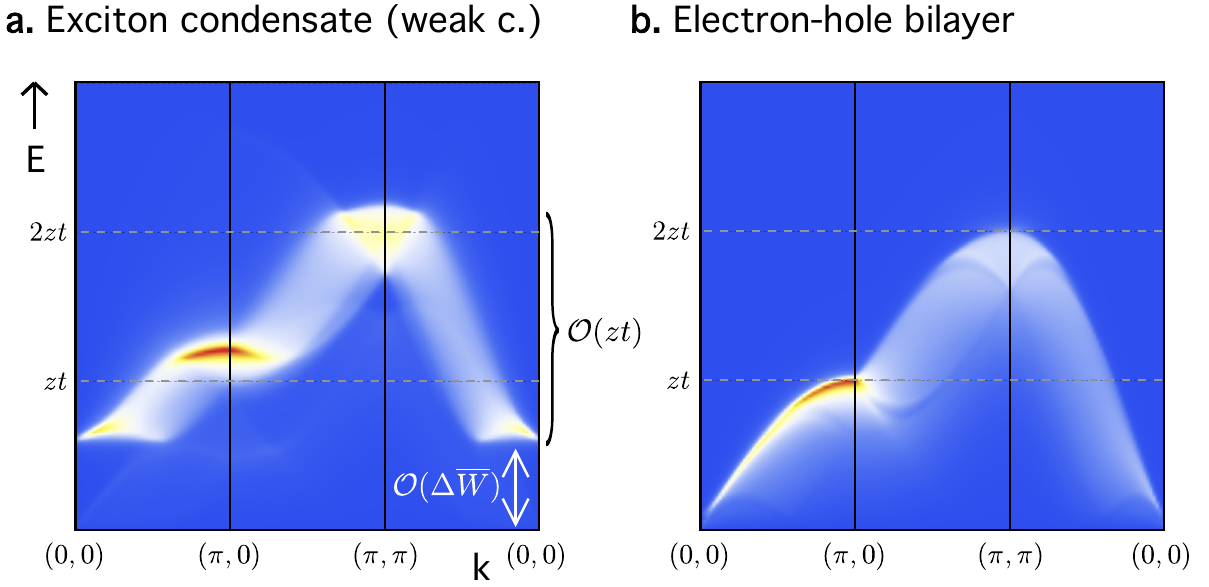}
	\caption{\label{FigWeakC} The absorptive part of the dynamical magnetic susceptibility $\chi''(\mathbf{q},\omega)$ in the weak-coupling limit of both the exciton binding energy and electron-electron interactions.
	\textbf{a:} The magnetic susceptibility is also in the exciton condensate phase dominated by the Lindhard continuum. This is qualitatively different from the triplons found in the strong coupling limit of figure \ref{FigStrongCLayerResults}.
	Model parameters are $\xi_{1\mathbf{k}} = - z t \gamma_\mathbf{k} - \mu = -\xi_{2\mathbf{k}}$, $t_\perp = 0.05 zt$, $\mu = -0.8zt$ and $\Delta \overline{W} = t$.
	\textbf{b:} For comparison we computed the $\chi''(\mathbf{q},\omega)$ in an electron-hole bilayer without exciton condensation.}
\end{figure}

Since the exciton condensate ground state is independent of the interaction strength, one can in principle adiabatically continue the strong coupling results to the weak coupling limit. However, in this limit the magnetic susceptibility as shown in figure \ref{FigWeakC} has a fundamentally different origin. Only with strong interactions the electrons are localized and a true spin degree of freedom arises. This is not the case for weak coupling, where the spin response is still dominated by the Lindhard continuum. The propagation scale of the triplet excitations is now set just by the dispersion of the noninteracting electrons.

To illustrate this point we compute the dynamic magnetic susceptibility for the weak coupling case where we depart from a band structure of electrons and holes $	H_K = \sum_{\mathbf{k}\sigma}\left(
		\xi_{1 \mathbf{k}} c^\dagger_{1 \mathbf{k}\sigma} c_{1 \mathbf{k}\sigma}
		+ \xi_{2 \mathbf{k}} c^\dagger_{2 \mathbf{k}\sigma} c_{2 \mathbf{k}\sigma} \right)$ plus a weak interlayer tunneling $H_\perp =  - t_\perp \sum_{\mathbf{k}\sigma} 
		\left( c^\dagger_{1 \mathbf{k}\sigma} c_{2 \mathbf{k}\sigma} 
			+ c^\dagger_{2 \mathbf{k}\sigma} c_{1 \mathbf{k}\sigma} \right)$
where $\xi_{\ell \mathbf{k}}$ is the band structure of the holes or electrons, depending on the layer. For simplicity, we take $\xi_{1\mathbf{k}} = - z t \gamma_\mathbf{k} - \mu = -\xi_{2\mathbf{k}}$ on a square lattice, so that in both layers there is an equal sized Fermi surface with opposite Fermi velocities. The interlayer hopping $t_\perp \ll t$ is assumed to be small given the insulator in between the layers. Both in-plane and the interlayer interactions are given by the Coulomb interaction
$H_V = 
		\sum_{ij \ell \sigma \sigma'} V_{ij} n_{i \ell \sigma} n_{j \ell \sigma'}
		+ \sum_{ij \sigma \sigma'} W_{ij} n_{i 1 \sigma} n_{j 2 \sigma'}
$, where $V_{ij} \propto |r_i - r_j|^{-1}$ and the interlayer Coulomb includes the interlayer distance $d$, hence  $W_{ij} \propto \left((r_i - r_j)^2 + d^2 \right)^{-1/2}$. The effects of these interactions are taken into account using the random phase approximation (RPA)\cite{BruusFlensberg}, in contrast to the linear spin wave approximation used in the strong coupling computation. In the bilayer case, one needs to extend the usual RPA expression $\chi = \chi_0 / (1 - V_q \chi_0)$ to include both intra- and interlayer interactions and bare susceptibilities $\chi_0$.

At some critical temperature the electron-hole bilayer has an instability towards exciton condensation. Based on the standard BCS theory \cite{Bardeen:1957p5048,DeGennesBook} we single out the interactions responsible for the singlet exciton pairing and perform a standard mean field decoupling using our earlier order parameter ansatz (\ref{OrderParameter}). This amounts to adding
$
	\sum_{\mathbf{k} \mathbf{k}'\sigma} W_{\mathbf{k}-\mathbf{k}'}
		\left( \Delta_{\mathbf{k}'} \Delta_{\mathbf{k}}^*
		 -  \Delta_{\mathbf{k}'} c^\dagger_{1 \mathbf{k}\sigma} c_{2 \mathbf{k}\sigma} 
			-  \Delta_{\mathbf{k}}^* c^\dagger_{2 \mathbf{k}\sigma} c_{1 \mathbf{k}\sigma} 
		\right) 
$
to the free Hamiltonian. Under the assumption that $\Delta_{\mathbf{k}}$ is independent of momentum it follows that condensation just amounts to an increase of the interlayer tunneling $t_\perp$. For any attractive momentum-averaged screened interaction $\overline{W}= \frac{1}{N} \sum_{\mathbf{k}} W_{\mathbf{k}}$ a condensate solution is found\cite{Shevchenko:1976p4950,Lozovik:2008p4877}. 

Let us fix the order parameter at a value of, say, $\Delta \overline{W} = t$. Using the aforementioned RPA expansion we compute the resulting magnetic excitation spectrum shown in figure \ref{FigWeakC}a. This spectrum is reminiscent of our strong coupling results of figure \ref{FigStrongCLayerResults}. But instead of the renormalization of the triplet bandwidth, the magnetic excitations closely follow the Bogolyubov quasiparticle spectrum. In fact, the dynamic magnetic susceptibility in the weak coupling limit can be best understood as a gapped variation of the result in absence of a condensate, shown in figure \ref{FigWeakC}b. In weak coupling, the gross features of the magnetic excitation spectrum therefore look similar with or without the exciton condensate, whereas the dramatic increase of the overall energy scale of the magnetic excitations is only present in the strong correlations limit.

In conclusion, we have shown explicitly that in a Mott bilayer the bandwidth of the magnetic excitations is strongly enhanced by the presence of an exciton condensate. We emphasize that this dynamic enhancement is quite unusual: the interplay between magnetic and charge degrees of freedom most commonly leads to frustration effects such as found in the $t-J$ model\cite{SingleHoleRefs,SingleHoleRefs2,Rademaker2012EPL}. Paradoxically, this effect turns around dealing with excitons in Mott insulators under the condition that they condense. This can promote the propagation of spin.

\emph{Acknowledgements ---}
This research was supported by the Dutch NWO foundation through a VICI grant. The authors thank Kai Wu for helpful discussions.

\end{document}